# Nonlinear optical microscopy with an obscuration-free, freeform reflective objective

YRYX Y. LUNA PALACIOS,[1] TUYET N. A. HOANG,[1] SALILE KHANDANI,[2] STEPHAN CLARK,[3] AARON BAUER,[4,5] JANNICK P. ROLLAND[4,5,6], ERIC O. POTMA[1,2], AND ADAM M. HANNINEN[7]

[1] *Department of Chemistry, University of California, Irvine, CA 92697, USA*
[2] *Department of Biomedical Engineering, University of California, Irvine, CA 92697, USA*
[3] *Clark Optical Consulting and Prototyping, Crowley, TX 76036, USA*
[4] *The Institute of Optics, University of Rochester, Rochester, NY 14627, USA*
[5] *Center for Freeform Optics, University of Rochester, Rochester, NY 14627, USA*
[6] *Department of Biomedical Engineering, University of Rochester, Rochester, NY 14627, USA*
[7] *Trestle Optics, Irvine, CA 92697, USA*

*\*epotma@uci.edu*

**Abstract:** Nonlinear optical (NLO) imaging platforms traditionally rely on refractive microscope objectives, which suffer from chromatic aberrations and temporal dispersion of pulsed excitation light. These issues degrade spatial imaging properties and signal brightness. Furthermore, the limited transmission range of refractive materials restricts NLO imaging, especially for applications requiring short- to mid-wave infrared excitation. While reflective microscope objectives offer an achromatic solution and broader transmission range (from visible to mid-infrared), conventional Schwarzschild designs have a central obscuration, which limits transmission throughput, imparts diffraction effects into the images, and, more generally, hinders the adoption of reflective optics in NLO microscopy. We overcome these obscuration-based limitations by developing a novel, reflective microscope objective using freeform mirrors in a non-coaxial geometry. This innovative, obscuration-free design boasts a 0.65 numerical aperture (NA), near diffraction-limited imaging performance, and offers significantly improved transmission with wider fields-of-view. We demonstrate its utility by integrating it into a standard laser-scanning microscope and performing NLO microscopy across a wide range of excitation wavelengths. Our freeform microscope objective outperforms standard reflective designs, providing an achromatic, dispersion-free alternative to refractive lenses for NLO imaging.

## 1. Introduction

Nonlinear optical microscopy comprises a suite of imaging techniques that generate image contrast from signals that scale nonlinearly with excitation intensity. The field gained prominence in the early 1990s with the advent of multiphoton excited fluorescence microscopy.[1-3] This breakthrough spurred the development of numerous other NLO modalities, such as second- and third-harmonic generation (SHG[4-7] and THG[8-10]), coherent Raman scattering (CRS)[11-14], and pump-probe imaging techniques.[15,16] Compared to linear optical microscopy, many multiphoton techniques exhibit reduced sensitivity to light scattering in biological tissues, enabling imaging at greater depths, a highly valuable feature for biological and biomedical imaging. Consequently, NLO imaging has become an indispensable microscopy method in laboratories globally.

In its most common form, NLO imaging is performed on a laser scanning microscope, whereby a tightly focused spot is raster scanned across the sample, and the targeted NLO signal is captured by a far-field detector. While NLO microscopes evolved from linear confocal

microscopes – sharing optical elements like scan/tube relay lenses and refractive microscope objectives[17] – their performance is fundamentally limited by these inherited refractive optics.

First, NLO microscopy relies on pulsed excitation light; shorter pulses incident on the sample yield brighter signals. However, chromatic dispersion within refractive elements causes color dispersion, which broadens pulse durations and diminishes NLO signal strengths.[18,19] While pulse pre-compression techniques offer partial mitigation[20-22], managing the shortest optical pulses remains challenging, and, in many cases, infeasible.

Second, refractive optics inherently suffer from spatial chromatic aberrations due to their wavelength-dependent refractive index.[23-25] Reducing these aberrations requires strategic combinations of flint and crown glass substrates, which increases the effective thickness of lens assemblies and consequently adds to the overall color dispersion of the microscope system. While axial and lateral chromatic aberrations over a finite wavelength range can be largely controlled in apochromatic designs, residual chromatic aberrations remain, which negatively affect NLO imaging properties. For instance, in multi-color techniques like CRS, they induce position-dependent misalignments across the field-of-view (FOV), resulting in signal degradation away from the FOV center. Furthermore, chromatic aberrations prevent the focusing of all spectral components of ultrabroadband laser pulses − critical for many NLO applications − to a single diffraction-limited spot, thus reducing signal strength as well as imaging resolution.[26]

Third, refractive optical components, and microscope objectives in particular, significantly limit the imaging system's transmission window. While common refractive microscope objectives typically transmit well between 0.4 μm and 1.2 μm, their throughput rapidly deteriorates beyond 1.4 μm. Although specialized objectives can extend the throughput to 1.7 μm, a crucial range for deep tissue three-photon excited fluorescence microscopy, fundamental material properties of optical glasses and complex anti-reflection thin film coatings make further expansion to longer wavelengths exceedingly difficult. Moreover, NLO imaging techniques like vibrationally sensitive sum-frequency generation (SFG)[27-30], fluorescence encoded infrared microscopy (FEIR)[31-33], and photothermal infrared (PTIR) imaging[34-36] require a combination of visible/near-infrared and mid-infrared excitation beams, which a single set of refractive optics cannot support.

In stark contrast, reflective microscope objectives offer inherent achromatism, negligible color dispersion, and a vastly expanded wavelength coverage from the visible to the mid-infrared. Despite these powerful attributes, their integration into laser scanning NLO imaging systems has remained notably limited[37-39], indicating a significant hurdle to widespread adoption. A primary drawback lies with the prevalent (and only commercially available) Schwarzschild design, whose on-axis secondary mirror blocks the center portion of the incident beam. While this concentric layout corrects spherical, astigmatic, and coma aberrations, in addition to chromatic aberrations, the central obstruction significantly reduces transmission throughput and produces a far-from-ideal point spread function (PSF). Furthermore, the Schwarzschild design cannot support a wide FOV, thereby limiting its practical utility for NLO imaging studies.

Bringing the inherent advantages of reflective optics to NLO microscopy necessitates overcoming the limitations of the common Schwarzschild reflective objective. To this end, off-axis designs offer a path forward. Yet, tilting curved reflective surfaces introduces complex aberrations that cannot be fully corrected by traditional rotationally symmetric mirror elements. This challenge has driven recent research towards incorporating freeform optical elements, which lack an axis of symmetry.[40] These unique surfaces provide the critical ability to compensate for aberrations beyond the scope of symmetric components.[41,42] For instance, one freeform design yielded an unobscured reflective objective with diffraction-limited performance (0.53 NA) across a wide 0.75 mm FOV[43], albeit at the cost of coaxial alignment of the image plane and its back focal (Fourier) plane. To resolve this issue, we recently engineered a compact, freeform reflective microscope objective[44] that boasts a 0.65 NA and a nominal diffraction-limited imaging performance across a 0.42 mm diameter FOV. Given its unobscured nature, combined with its compatibility with standard microscope nosepiece turrets, this bold optic holds significant promise for advancing NLO and multimodal imaging studies.

In this work, we demonstrate the new capabilities enabled by the first-of-its-kind unobscured, custom freeform reflective microscope objective developed for biological NLO microscopy. We experimentally validate the new 0.65 NA objective prototype across various NLO imaging modalities. A direct comparison to a commercial Schwarzschild reflective objective of similar NA reveals the distinct advantages of the freeform design, effectively removing existing hurdles to achieving dispersion-free NLO microscopy across the entire visible-to-mid-infrared spectral range.

## 2. Design concept and experimental validation

### 2.1 Optical and optomechanical design

The unobscured reflective microscope objective is based on a six-mirror design.[44] Briefly, the design includes a three-mirror front group that focuses collimated light to an intermediate image plane, then a fold mirror for light redirection, followed by a rear two-mirror group that demagnifies the intermediate image and forms the tight focus at the image plane, as shown in Figure 1(a). The combined layout incorporates three freeform mirrors, two spherical mirrors, and one flat fold mirror. The unobscured microscope objective is infinity corrected, and its entrance and exit apertures are coaxial, mimicking the inherent geometric in- and out-puts of traditional refractive objectives.

The optomechanical design includes a base plate onto which the mirror components are individually mounted, as illustrated in Figure 1(b). The mounting base is equipped with key reference datums that facilitate the precise placement of the mirror substrates with dowel pins. The mirror assembly is covered by an enclosure that includes the input and output apertures, as well as an RMS-threaded mounting base for placement onto a standard microscope turret. The resulting parfocal height is 70 mm, and its diameter measures 50 mm. The entrance aperture is fixed at 7 mm, and the usable working distance is 1 mm.

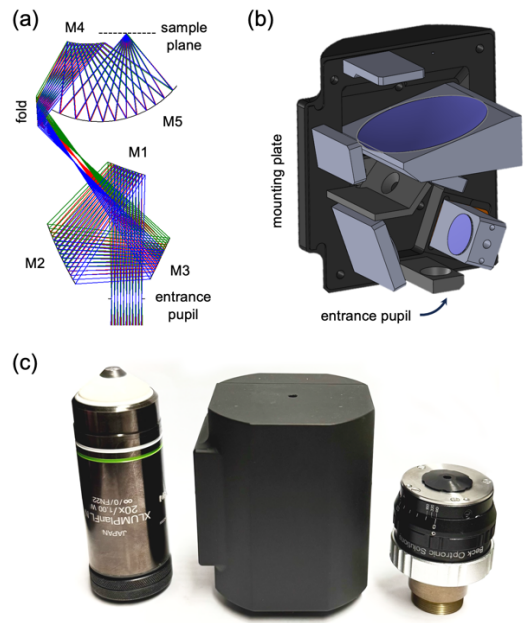

**Figure 1**. Freeform, unobscured reflective microscope objective. (a) Six mirror design, including two spherical mirrors (M1, M3), three freeform surfaces (M2, M4, M5), and one flat mirror. (b) Optomechanical design showing mirrors assembled on the mounting plate. (c) Size comparison between a 20x, 1.00 NA refractive objective (Olympus, XLUMPLFLN), the unobscured 0.65 NA reflective objective, and a 0.65 NA Schwarzschild objective (Beck, 5007).

### 2.2 Nonlinear optical microscopy experiments

To validate the assembled reflective microscope objective for NLO laser scanning microscopy, we incorporated it into two different optical imaging platforms. The first imaging system was based on a commercial laser scanning module (Fluoview 300, Olympus), interfaced with an inverted microscope frame (IX71, Olympus). Equipped with a 3 ps, 80 MHz beam at 800 nm from a parametric optical oscillator (picoEmerald, A.P.E. Berlin), this imaging system was used to collect two-photon excited fluorescence signals from polystyrene beads, in addition to linear transmission images. Signals were spectrally filtered and detected by a photomultiplier tube (R3896, Hamamatsu).

A second imaging system comprised a galvo scanner (LSKGG4, Thorlabs ), a custom-built reflective relay, and an inverted microscope frame (IX71, Olympus) to enable all-reflective imaging from the visible to the mid-infrared wavelength regimes.[39] A 5 ps, 76 MHz parametric optical oscillator (Levante IR, A.P.E. Berlin), synchronously pumped by a 1031 nm Yb-based laser (aeroPULSE, NKT Photonics), was used as the light source. NLO signals (SHG, SFG, coherent anti-Stokes Raman scattering) were generated in the collinear excitation geometry, collected with a refractive condenser in the forward direction, spectrally filtered, and registered by a pair of photomultiplier tubes (H16721-50 & H7422-01, Hamamatsu). For the imaging studies reported herein, all tissue samples were mounted on standard 0.17 mm coverslips before their inspection in the inverted microscope.

### 3. Results

#### 3.1 Field-of-view and resolution

A key design target of our reflective microscope objective was the expanded FOV compared to a Schwarzschild objective of similar NA, while maintaining a high lateral resolution. To establish the performance of the prototype reflective objective, we first examined its linear imaging properties using a USAF 1951 resolution test chart as the sample. Figure 2(a) shows the transmission image obtained by laser-scanning the 800 nm beam across the field-of-view. It can be seen that good image contrast and visibility are maintained over a FOV that extends to 0.4 mm, in agreement with our original design target. A line scan across element 6 of group 7 of the test chart (2.19 µm line width) is shown in Figure 2(b).

To assess the NLO imaging properties of the reflective microscope objective, we measured the two-photon excited fluorescence signal from 0.2 µm dye-loaded polystyrene beads suspended in an agarose matrix, excited at 800 nm. As shown in Figure 2(c), individual beads were clearly resolved. Figure 2(d) displays a lateral cross-section, obtained by statistically averaging measurements from ten beads. A Gaussian fit (red) to this data reveals a full width at half maximum (FWHM) of 0.68 µm. The red fitted curve is close to the theoretically predicted width for a diffraction-limited two-photon excitation profile of a 0.65 NA objective after convolution to account for the polystyrene bead's size (depicted as the grey curve). Together, these measurements show that the new reflective microscope objective affords a relatively wide FOV combined with sub-micrometer resolution.

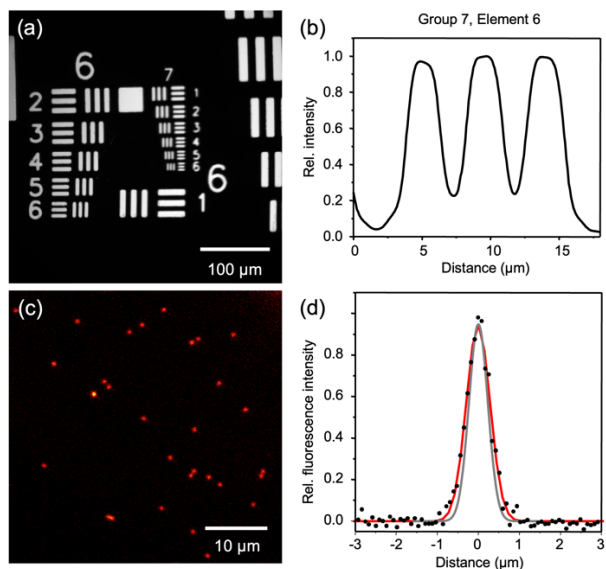

**Figure 2**. Field-of-view and lateral resolution assessment of the prototype reflective objective. (a) Linear transmission image of a USAF 1951 test chart, recorded at $\lambda = 800$ nm in laser-scanning mode. (b) A lateral cross-section of vertical element 6 of group 7. (c) Two-photon excited fluorescence of dye-loaded 0.2 µm polystyrene beads ($\lambda = 800$ nm). (d) Average of ten lateral bead profiles (black dots) and a Gaussian fit (red, FWHM. = 0.68 µm), along with the calculated diffraction-limited response convolved with the 0.2 µm bead size (grey).

*3.2 Achromatic imaging over an extended wavelength range*

Several emerging NLO imaging techniques, such as vibrationally-sensitive SFG, FEIR, and PTIR, rely on two excitation beams with widely separated wavelengths. Traditional refractive objective lenses do not support the preferred collinear excitation scheme due to their limited transmission window. While reflective microscope objectives address this, the low transmission and distorted PSF of Schwarzschild objectives have hindered wider adoption of these infrared-sensitive NLO methods. Here, we show that a freeform, unobscured reflective microscope objective overcomes the transmission limitations while also offering the relatively wide field of view needed for tissue imaging applications.

We used the newly developed 0.65 NA unobscured microscope objective to perform SHG and mid-infrared sensitive SFG imaging on rat tail tendon tissues, and directly compared its performance to a commercial 74x, 0.65 NA Schwarzschild objective (Beck, 5007).

A key issue with Schwarzschild objectives is their inherent light loss. Our measurements confirmed this: a 633 nm He-Ne laser beam (1/e width of 5 mm) experienced only 32% throughput with the Schwarzschild, largely due to aperture clipping and beam blocking by its central obscuration. In contrast, the unobscured freeform design achieved 78% throughput, proving its efficiency in minimizing losses. The remaining optical losses are mainly attributed to the accumulated reflectivity of the objective's six mirrors ($R \sim (0.97)^6 = 0.83$) rather than to vignetting losses.

Beyond throughput, the Schwarzschild objective also compromises image quality and field-of-view. As seen in the SHG image of the tendon's collagen type I (Figure 3a, 1031 nm pump), captured with the Schwarzschild, signal degradation at image margins limits the effective FOV. The unobscured freeform objective overcomes this degradation by supporting wider scan angles by reducing vignetting, yielding a significantly larger observable area of the tissue sample. Moreover, by transmitting the full angular distribution afforded by its 0.65 NA, the freeform objective exhibits a modulation transfer function (MTF) comparable to refractive objectives.[44] The superior spatial frequency transmission over the Schwarzschild reflective objective means images captured with the unobscured all-reflective freeform objective contain significantly more detail and spatial features. The improved performance also translates to a better photon budget; we acquired SHG images with our prototype using three times less input power than the conventional reflective objective. Lowering photon budgets can synergize with lower-cost lasers and, consequently, lower-cost microscopy.

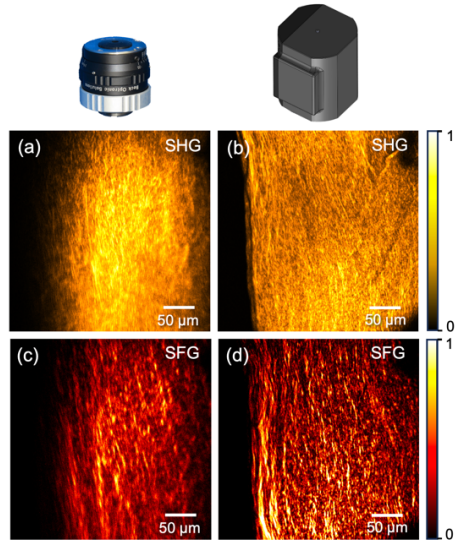

**Figure 3**. Comparison between the NLO imaging performance of a conventional Schwarzschild objective (0.65 NA) and the newly developed unobscured freeform objective. (a) SHG image of rat tail tendon acquired with the Schwarzschild objective ($\lambda_{ex}$ = 1031 nm), and (b) the same image obtained by the freeform objective. SFG images ($\lambda_{MIR}$ = 3400 nm, $\lambda_{NIR}$ = 1031 nm) of the same area, acquired by the Schwarzschild objective (c) and the reflective freeform objective (d).

The same advantages of the unobscured freeform design are evident in the SFG images obtained from the same sample, as shown in Figures 3(a) and 3(b). For these experiments, the mid-infrared beam was tuned to 3400 nm (2941 cm$^{-1}$) to drive the carbon-hydrogen vibrational stretching mode of collagen's methylene groups, while the upconversion beam was fixed at 1031 nm to generate an SFG signal at 791 nm. The improved imaging performance, both in terms of image features, FOV, and reduced excitation power, allows a more thorough examination of the tissue sample over expanded areas. This is further exemplified in Figure 4, which shows a mosaic composed of 0.3 x 0.3 mm$^2$ tiles, offering a mesoscale view of the tendon tissue. Such mesoscale images are significantly more time-consuming to acquire with the Schwarzschild objective, whose FOV diameter is more than three times smaller (> 9 times smaller area).

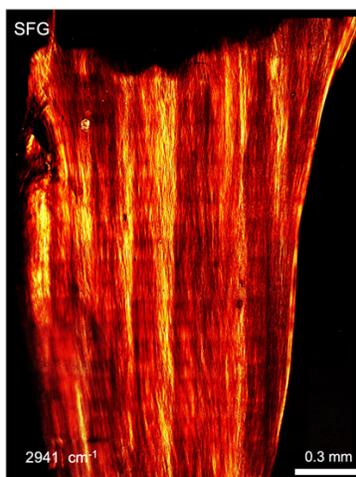

**Figure 4**. Mesoscale SFG map of tendon tissue composed of 0.3 x 0.3 mm$^2$ tiles acquired with the freeform objective. The mid-infrared beam is tuned to the 2941 cm$^{-1}$ CH$_2$ vibrational resonance of collagen Type I.

### *3.3 Multimodal NLO microscopy*

The new unobscured all-reflective freeform objective enhances NLO imaging by offering throughput, field-of-view, and image quality comparable to refractive objectives, but critically, extending these benefits deep into the mid-infrared range. This expanded wavelength window enables uncompromised imaging with a broader selection of NLO contrast mechanisms on a single microscope platform. To illustrate this versatility, we performed NLO tissue imaging across various excitation wavelengths and modalities.

As shown in Figure 5(a), we obtained an SHG image with a 1031 nm excitation wavelength, while Figure 5(b) displays an SHG image acquired at 1570 nm. Both wavelengths are fully supported by the unobscured reflective freeform objective, incurring no additional transmission losses thanks to the nearly constant reflectivity of protected silver from 0.4 µm to 10 µm.

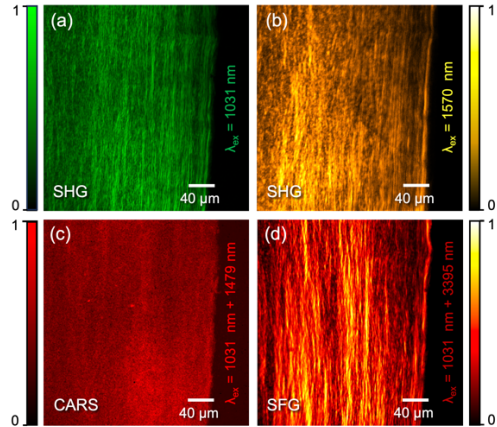

**Figure 5**. Multimodal NLO imaging with an unobscured all-reflective freeform objective. SHG imaging of collagen type I at $\lambda_{ex}$ = 1031 nm (a) and $\lambda_{ex}$ = 1570 nm (b). Vibrationally sensitive imaging of the $CH_2$ stretching mode at 2946 cm$^{-1}$ with CARS (c) and SFG (d) microscopy.

Further demonstrating its capabilities, Figures 5(c) and 5(d) present vibrationally sensitive images where contrast is derived from the 2946 cm$^{-1}$ $CH_2$ stretching mode of the collagen-rich tissue. Figure 5(c) shows a coherent anti-Stokes Raman scattering (CARS) image, generated using a 1031 nm pump beam and a Stokes beam tuned to 1479 nm. In contrast, Figure 5(d) depicts an SFG image, probing the same vibrational mode with a mid-infrared excitation beam at 3395 nm combined with a 1031 nm upconversion beam. The broadband, achromatic performance of the all-reflective freeform objective allows both CARS and SFG signals to be registered on the same system using a straightforward collinear excitation scheme, which is ideal for multimodal laser-scanning microscopy. This unique capability facilitates a direct comparison between the vibrationally sensitive SFG image and the corresponding CARS image, revealing distinct spatial information about the tissue.

## 4.  Discussion

Reflective imaging systems are, in principle, superior to refractive designs due to their inherent achromaticity and broad spectral transmission. Yet, the scarcity of commercial or custom-built reflective laser-scanning optical microscopes highlights the significant limitations of traditional Schwarzschild microscope objectives. Their limited throughput and non-ideal focusing qualities have genuinely stifled advancements in nonlinear optical (NLO) imaging, especially for techniques using excitation wavelengths beyond 1.7 µm that critically depend on high-performance reflective optics.

As demonstrated by our imaging experiments, the development of an unobscured reflective microscope objective overcomes many of the Schwarzschild objective's drawbacks. The compact, all-reflective solution presented in this work is made possible by freeform design concepts, which allow the creation of this type of unobscured folded optics. This innovation opens up new capabilities for achromatic NLO microscopy across an expansive 0.4 µm to 10 µm wavelength range. The newly developed 0.65 NA prototype objective delivers spatial resolutions close to the theoretical limit, alongside enhanced throughput, a well-behaved MTF, and a wide FOV when compared to a conventional Schwarzschild objective of similar NA. This effectively eliminates many long-standing disadvantages associated with reflective objectives that have a central obscuration. Furthermore, because our prototype's form factor was optimized for integration with standard microscope frames, these improved imaging attributes are directly applicable to NLO imaging studies in common research settings.

While the current prototype clearly demonstrates the benefits of an unobscured reflective objective freeform design, we anticipate further improvements will solidify its position as a comprehensive imaging solution, potentially replacing standard refractive objectives in nonlinear optical (NLO) microscope platforms.

We note that despite impressive advancements in manufacturing freeform reflective surfaces over the past decade, their fabrication remains significantly more challenging than that of rotationally symmetric optics, especially if manufacturers must rely on costly computer-generated holograms (CGHs) to achieve the form of each freeform surface.[45] While CGHs are the standard metrology tool for volume production, emerging technology in reconfigurable CGHs is poised to permeate the manufacturing of freeform optical elements for limited series and early prototyping.[46,47] The lack of readily available freeform metrology tools at most diamond machining facilities makes real-time feedback during the diamond machining process unfeasible, which compromises fabrication accuracy and tolerance limits. For example, metrology of our prototype's freeform mirrors revealed out-of-spec surface profiles that impacted the overall performance.[48] Specifically, we found that residual astigmatic aberrations induced by small assembly misalignments broadened the current unit's focal volume, particularly along the axial dimension. These fabrication deficiencies can be eliminated by integrating a more streamlined, metrology-enabled feedback loop during production, ensuring the microscope objectives' as-built performance aligns more closely with their nominally designed performance.

Furthermore, many NLO imaging studies require properties that exceed our prototype's current specifications, such as an even higher NA, a wider FOV, an increased working distance, or combinations of these. The success of our unobscured reflective objective suggests that future enhancements are feasible. For instance, our measured FOV of 0.4 mm diameter suggests that, with additional design adjustments, it is possible to extend this metric well beyond 0.5 mm. While achieving a higher NA and longer working distances presents greater challenges due to geometrical constraints, alternative designs that leverage liquid immersions are expected to push for these improvements. Considering that refractive optics have benefited from centuries of optimization, it is reasonable to expect that a similar evolution in the relatively young field of freeform optics will lead to significant advancements in reflective objective design.

Ultimately, the results presented here establish the compelling potential for all-reflective unobscured imaging, pointing toward a future of fully achromatic, broadband, high-resolution NLO imaging.

## 5. Conclusion

In this work, we developed a first-of-its-kind unobscured all-reflective microscope objective, enabled by freeform optics, and demonstrated its transformative capabilities for NLO microscopy. The 0.65 NA microscope objective features a 0.4 mm diameter FOV, a lateral resolution close to the theoretical limit, and a high throughput of 78% across the expansive 0.4 μm to 10 μm spectral range with perfect color correction. Imaging examples of tendon tissue, using the SHG, CARS, and mid-infrared sensitive SFG imaging modalities, revealed the objective's superior imaging properties compared to a conventional Schwarzschild objective of identical numerical aperture. By overcoming long-standing limitations, these results unequivocally highlight the renewed potential of reflective optics in NLO microscopy, enabling truly achromatic imaging from the visible to the mid-infrared spectrum.

**Funding.** National Institutes of Health (R43GM149018, R21-EB034084); National Science Foundation (2221721, 2404006).